\documentclass[12pt]{article}
\usepackage{graphicx}

\def\beq{\begin{equation}}
\def\eeq{\end{equation}}
\def\bfg{\begin{figure}}
\def\efg{\end{figure}}
\def\cpt #1{ \parindent1in\parbox{4.5in}{\caption{\footnotesize #1}} }
\def\address #1{ \begin{center}\small\it #1 \end{center} }

\title{Transparent photonic band in metallodielectric nanostructures}
\author{Simin Feng,  J. Merle Elson,  and Pamela L. Overfelt}

\begin{document}

\maketitle
\address{Naval Air Warfare Center, Research Group \\ Physics and Computational Sciences, China Lake, CA 93555}

\abstract{\baselineskip=2.5\baselineskip{
Under certain conditions, a transparent photonic band can be designed into a one-dimensional metallodielectric nanofilm structure.  Unlike conventional pass bands in photonic crystals, where the finite thickness of the structure affects the transmission of electromagnetic fields having frequency within the pass band, the properties of the transparent band are almost unaffected by the finite thickness of the structure.  In other words, an incident field at a frequency within the transparent band exhibits 100\% transmission independent of the number of periods of the structure.  The transparent photonic band corresponds to excitation of pure eigenstate modes across the entire Bloch band in structures possessing mirror symmetry.  The conditions to create these modes and thereby to lead to a totally transparent band phenomenon are discussed.  \\

\vskip-.3in


\pagebreak

\parindent.3in
\section{Introduction}
Propagation of electromagnetic waves through periodic structures has both fundamental interest and potential applications because of the possibility of controlling optical properties of materials through photonic band gaps\cite{Yablon}--\cite{Bendickson}.  Strictly speaking, a band diagram describes properties of an infinite periodic structure.  In finite structures, there are transmission resonances inside pass bands\cite{Bendickson}, i.e. minibands separated by minigaps as demonstrated by the experiment\cite{Olivier}.  The bandwidth and central frequency of each resonance change with the number of the periods.  The finite thickness or termination of photonic crystals (PhCs) has a significant impact on the dispersive properties of the structure\cite{Ye}--\cite{Elson} and hence affects transmission properties.  In particular, the transmission coefficient of a plane wave having frequency within the pass band exhibits a thickness-dependent periodic oscillation.  This thickness-dependent transmission oscillation and length-sensitive behavior of the PhCs can impact many potential applications and photonics integrations.  In this paper, we analyze transmission features of photonic bands and show that under certain conditions a transparent photonic band can be formed in one-dimensional metallodielectric nanofilms.  The term, transparent photonic band, refers to two significant differences from the conventional pass bands.  First, there are no minigaps and minibands inside the transparent band in a finite structure.  Second, the electromagnetic fields having frequency within the transparent band exhibit complete transmission independent of the number of periods of the structure.  The properties of the transparent band are almost unaffected by the finite thickness of the structure and this is the major difference when compared to the conventional pass bands.  The existence of the transparent band depends on the surface plasmon excitation and the manner in which the first and last layers are truncated.  More precisely, the transparent band corresponds to pure eigenstate excitations across the entire Bloch band in structures possessing mirror symmetry.  Each frequency in the transparent band is in an eigenstate common to translation and surface-wave operators.  In the transparent band, the electric field intensity is symmetric across the length of the structure.  The mirror symmetry of the physical structure is essential for the existence of the transparent photonic band.  The conditions to create this totally transparent phenomenon will be discussed in the following sections.  In this regard, we cite an earlier experiment exploring the possibility of rendering a layered metallic structure to be transparent\cite{Scalora}.

\section{ Coupled plasmon resonant waveguide (CPRW) }
The investigation is performed with a 1D structure that has alternating layers of metal and dielectric materials.  The essential physics of the 1D structures should be equally applicable to higher dimensional structures while avoiding intensive computational tasks of higher dimensionality.  Also, the 1D structure is relatively easy to fabricate.  In the direction perpendicular to the layers, the waveguiding is achieved through evanescent coupling of surface plasmons between interfaces of metal/dielectric layers, i.e. a coupled-plasmon-resonant waveguide (CPRW).  The CPRW described here is a special case of the coupled-resonator optical waveguide (CROW) proposed by Yariv {\it et al}\cite{Yariv}.  Differing from the conventional PhCs, the band structures in the CROWs are formed from localized resonant modes through evanescent coupling.  Specifically, for the CPRW structure presented here, the evanescent coupling is achieved through a coupled chain of surface plasmon excitations at each interface, conveying the fields from layer to layer throughout the entire structure.  Meanwhile, the surface plasmons provide a large field enhancement at the interfaces of metal/dielectric layers.  The metallodielectric structures studied previously in the context of PhCs\cite{Kuzmiak}--\cite{Fan} where the Bloch modes of traveling waves were considered.  Here we will explore this type of structure from the perspective of the CROW and focus on the Bloch modes of the evanescent waves.  In our case, the fields inside each layer are evanescent along the propagation direction of the Bloch waves, i.e. the direction perpendicular to the interfaces of the layers.  In the CPRW structure, the waveguiding is achieved simultaneously parallel to the interfaces and perpendicular to the interfaces with different guiding mechanisms.  Parallel to the interfaces, the electromagnetic fields are guided by the surface plasmons whereas perpendicular to the interfaces the waveguiding is achieved through the evanescent coupling.  In such structures, we focus on the scenario that the field is evanescent inside all the layers, yet, a 100\% transmission can still be achieved for infinitely thick composite materials.  To achieve this goal, the nature of the surface plasmons is well suited to periodically amplify the evanescent fields at each metal-dielectric interface.  Due to the surface plasmon resonance and large field amplitudes at the interfaces, the termination of the CPRW structure can have a dramatic effect on the transmission  and this can be more pronounced than in the conventional photonic crystals.

\section{Theory}
Figure~\ref{StructCPRW} shows a one-dimensional lossless metallodielectric multilayer structure.  The period is $d=d_1+d_2$ where $d_1$ and $d_2$ are the thicknesses of the dielectric and metallic layers.  The subscript 1 refers to the dielectric medium with non-dispersive permittivity $(\epsilon_1)$ and permeability $(\mu_1)$.  The subscript 2 refers to the metallic material with constant $\mu_2$, and dispersive $\epsilon_2(\omega)=1-\omega_p^2/\omega^2$ where $\omega_p$ is the bulk plasma frequency.  In general, the $\omega_p$ is a function of not only electron density but also surface structure, such as subwavelength holes or slits.  We assume the surface plasmon  waves propagate along the $\bf\hat x$ direction with wave number $K_p$ and consider TM polarization with ${\bf E}=(E_x,0,E_z)$.  The electric field in the {\it n}th unit cell\cite{Simin} is given by 
${\bf E_n}({\bf r})={\bf \mathcal E}_{\bf n}(z)\exp\left(iK_px\right)$, where the evanescent portion of the fields are
\begin{eqnarray}
\label{E}
{\bf \mathcal E}_{\bf n}(z) = \cases{
{\bf a_n} \exp\bigl[-\alpha_1 z_n \bigr] + 
{\bf b_n} \exp\bigl[\alpha_1 (z_n-d_1) \bigr], 
& $0 \le z_n < d_1$  \cr  \noalign{\medskip}
{\bf c_n} \exp\bigl[-\alpha_2(z_n-d_1) \bigr] + 
{\bf d_n} \exp\bigl[\alpha_2 (z_n-d) \bigr], 
& $d_1 < z_n \le d$ } \ .
\end{eqnarray}
where $z_n = z - nd$ and $\alpha_i = \sqrt{K_p^2 - k_0^2\epsilon_i\mu_i} > 0,\,i=1,2$.
The field amplitudes $\bf a_n, b_n, c_n$, and $\bf d_n$ are vectors in the $\bf\hat x$--$\bf\hat z$ plane.  The relationship between these amplitudes can be found by enforcing the boundary conditions and the fact that $\nabla\cdot{\bf E}=0$ inside each layer.

Using the Bloch theorem, the evanescent condition $K_p^2>k_0^2\epsilon_1\mu_1$, and the material dispersion $\epsilon_2(\omega)=1-(\omega_p/\omega)^2$, with $\omega<0.6\,\omega_p$, the band structure of the coupled evanescent fields is obtained and these results are shown in Fig.~\ref{HMband} for different layer thicknesses.  The blue zones represent the pass bands of the evanescent fields, i.e. the Bloch modes of the evanescent fields.  For each period, the mode amplitudes are symmetric in the lower frequency band and anti-symmetric in the upper band.  These modes possess features of both surface plasmons and Bloch waves.  Both transmission pass bands are below the light line of the dielectric medium and this means that the fields are evanescent not only in the metal layers but also in the dielectric layers.  The upper band is truncated at the light line so that the modes are evanescent and bounded to the metal/dielectric interfaces.  In this paper, we only consider resonant transmission of bounded surface modes.  In this scenario, all the layers must be thin enough to allow overlap of the evanescent fields associated with the bounded surface modes.  Notice that the field decays exponentially inside each layer and is amplified at each interface.  If any layer is too thick such that the amplification at the interface cannot overcome the exponential decay inside the layer, there will be no transmission.  However, the total thickness of the structure can be infinite provided that each layer is sufficiently thin to allow coupled eigenstates throughout the entire structure.  Hence, the periodic amplification by the surface plasmon at each interface delivers the evanescent field from the first layer of the structure to the last layer.

\section{Numerical results}
The band diagrams in Fig.~\ref{HMband} describe the transmission properties of an infinitely periodic structure.  Since a practical structure will have a finite number of periods, the transmission properties are generally dependent on the thickness of the structure.  The transmission coefficient of the finite structure can be calculated using a transfer matrix method.  The incident plane wave is coupled with the Bloch evanescent modes via superstrate and substrate coupling prisms and this allows excitation of the surface plasmon field.

Figure~\ref{HMtransSpec} shows the evolution of the photonic bands with increasing the number of periods of the structure.  The plots on the left and right sides correspond to different surface plasmon wave numbers where $K_p=2.7\pi/d$ on the left and $K_p=1.6\pi/d$ on the right.  The layer thicknesses $d_1$ and $d_2$ in Fig.~\ref{HMtransSpec} are the same as those in Fig.~\ref{HMband}(b).  There are two pass bands associated with each surface plasmon excitation.  The correlation between the band diagrams and transmission spectra can be seen by comparing Fig.~\ref{HMtransSpec}(c) and (f) with Fig.~\ref{HMband}(b).  
In Fig.~\ref{HMtransSpec}(c) and (f), there are four transmission pass bands centered about $820$, $855$, $625$ and $725$\,THz. The $820$, $625$ and $725$\,THz bands are not transparent bands and the rapid oscillations within these pass bands are mini-resonances.  The transmission coefficient is very sensitive to the frequency.  The band centered about $855$\,THz in Fig.~\ref{HMtransSpec}(c) is the transparent photonic band.  This transparent band has a flat top spectrum and is almost immune from the rapid oscillation.  The bandwidth of the transparent band is almost unchanged after the first few periods.

The striking feature of the transparent band is that it allows 100\% transmission independent of the thickness of the structure.  This thickness-independent transmission is illustrated in Fig.~\ref{HMtransCoef}(a) and this feature is due to the single mode (pure eigenstate) character of the transparent band.  As a comparison, Fig.~\ref{HMtransCoef}(b) and (c) show the thickness-dependent transmission of the non-transparent pass bands and that the transmission coefficients periodically oscillate with the number of periods of the structure.  Finally, Fig.~\ref{HMtransCoef}(d) describes the character of the stop band where the transmission decays exponentially with increasing the number of the periods.

To better understand the transparent phenomenon, Fig.~\ref{HMkvects} shows the evolution of Bloch wavevector $K_B$ versus the number of the periods.  The frequencies in these plots correspond to those in Fig.~\ref{HMtransCoef}.  For the stop band in Fig.~\ref{HMkvects}(d) the value of $K_B$ decays exponentially with the number of the periods and eventually becomes imaginary.  In the transparent band Fig.~\ref{HMkvects}(a), the Bloch wavevector is constant and satisfies the dispersion of the infinite structure.  So, each frequency in the transparent band is in a pure eigenstate of the translation operator, and the field resembles a plane wave with a single $K_B$ for each frequency.  The propagation direction is given by the surface plasmon and Bloch wavevectors $({\bf K_p}, {\bf K_B})$.  For the other two cases, $K_B$ oscillates with a large amplitude for the first few periods, then it converges to oscillate about the $K_B$ of the infinite structure with a small amplitude as shown in plots Fig.~\ref{HMkvects}(b) and (c).  The spread of the $K_B$ is much less than the size of the Brillouin zone, i.e. $\Delta K_B\ll2\pi/d$.  The k-plots explain the oscillation and wave-packet phenomenon in the transmission coefficients.  The carrier of the oscillation is determined by the $K_B$ of the infinite structure, whereas the envelope of the oscillation is inversely proportional to the bandwidth of the $K_B$.  The stronger the coupling, the wider the spatial extent of the transmission coefficient, and hence the narrower the bandwidth in the k-space.  Thus, unlike the transparent band, each frequency in the conventional pass bands is in a superposition state, i.e. a superposition of the Bloch modes with the major component given by the corresponding infinite structure.  Notice that the positions of the nodes in Figs.~\ref{HMkvects}(b) are the same as the positions of the maximum transmission in Fig.~\ref{HMtransCoef}(b).  The 100\% transmission happens when the Bloch wavevector matches that of the corresponding infinite structure.  The number of the periods ($N$) between the peaks satisfies the relation $K_BNd\approx m2\pi$, where $m$ is an integer.

\section{Unit cell configuration}
The unit cell configuration is very important to achieve the transparent photonic band.  This is because the unit cell determines how the metal-dielectric structure is physically terminated.  First, the unit cell $M_{1/2}DM_{1/2}$ or $D_{1/2}MD_{1/2}$ can yield a transparent band, but the $MD$ configuration does not.  Note that for a metal-dielectric structure made up of $N$ unit cells,  the cases $(M_{1/2}DM_{1/2})^N$ and $(D_{1/2}MD_{1/2})^N$ are terminated identically at the both ends whereas the case $(MD)^N$ is not.  Analyzing the field distribution inside the structure can provide physical insight.  Figure~\ref{HMfield} shows the intensity distribution inside an isolated unit cell $M_{1/2}DM_{1/2}$.  There is naturally a field enhancement at the interfaces of the metal/dielectric layers due to the bounded surface waves.  The case for the transparent band is shown in Fig.~\ref{HMfield}(a), where the intensity distribution is symmetric.  In Fig.~\ref{HMfield}(b) and (c), the fields are not symmetric and are outside the transparent band.  Here is the key point: the intensity is symmetric in the unit cell when the transparent photonic band is excited.  Thus, each frequency in the transparent band is in an eigenstate of the surface-wave operator since the symmetric and anti-symmetric fields are the eigenmodes of the surface waves.  It is evident that when the unit cells that have mirror symmetry are layered together, the resulting modes can fit easily with a minimal adjustment.  On the other hand, if the non-symmetric $MD$ cell is layered, the modes do not fit together and this results in large field discontinuities and mode splittings.  The fields of the two resonant peaks in Fig.~\ref{HMtransSpec}(d) also exhibit unit transmission regardless of the length of the structure, and the field amplitude distribution is symmetric for the lower frequency and anti-symmetric for the higher frequency.  However, they are isolated frequencies without the bandwidth, unlike the case in Fig.~\ref{HMtransSpec}(c) that the entire Bloch band is excited into the transparent band.

Shown in Fig.~\ref{HDMtrans} is a comparison of the transmission spectra of the three configurations, $(MD)^N$, $(M_{1/2}DM_{1/2})^N$, and $(D_{1/2}MD_{1/2})^N$.  For the same constituent materials and layer thicknesses, the excitation condition for the transparent photonic band is different for the configurations $(M_{1/2}DM_{1/2})^N$ and $(D_{1/2}MD_{1/2})^N$.  Also, the transparent band is excited in the upper band for the $(M_{1/2}DM_{1/2})^N$ and in the lower band for the $(D_{1/2}MD_{1/2})^N$.  In contrast with the unit cells $M_{1/2}DM_{1/2}$ and $D_{1/2}MD_{1/2}$ where the bandwidth of the transparent band is almost unchanged upon increasing the number of the periods, the mode of the unit cell $MD$ cannot survive.  When two $MD$ cells are brought together the mode is split as shown in Fig.~\ref{HDMtrans}(a) and (b).  On the other hand, when the $(M_{1/2}DM_{1/2})^N$ or $(D_{1/2}MD_{1/2})^N$ cells are brought together, the pass band splits into a transparent band and a non-transparent band.  In the configurations $(M_{1/2}DM_{1/2})^N$ or $(D_{1/2}MD_{1/2})^N$, the two end surfaces are identical and that, under the proper excitation condition, can lead to a strong coupling of the two surface modes throughout a series of plasmon excitations at the interior metal-dielectric interfaces.  Hence, the unit cells of the metallodielectric periodical structures can have a great impact on the transmission properties.  Since the transparent band corresponds to the excitation of the entire Bloch band, the bandwidth of the transparent band can be up to 100\,THz with a flat-top spectrum depending on the thickness of the layers.

Shown in Fig.~\ref{HMdisper} are the transmission spectrum, dispersion, and group velocity of a 5-period CPRW.  In the dispersion plot, the four lines of small slope in the lower band contain dispersion signature of the CPRW in a weak coupling limit and may be useful in a light stopping scheme\cite{Yanik}.  The large slope line sitting on the upper branch of the dispersion curves of the infinite structure is the dispersion of the transparent band.  It occupies the entire positive k-space in the first Brillouin zone.  This indeed confirms that the transparent band is an excitation of the entire Bloch band.

\section{Conclusions}
We have analyzed the transmission features of photonic bands in the one-dimensional all-evanescent metallodielectric structure.  The transparent photonic band can be excited in the finite structure provided that the structure is composed with unit cells having mirror symmetry.  This has the crucial property of allowing very efficient coupling of an incident beam into the surface plasmon modes of the metal-dielectric structure.  With this and the proper excitation condition, the transmission from the input side to the output side of the metal-dielectric structure can be 100\% independent of the number of periods of the structure.  The transparent photonic band may also exist in other CROW structures.  Since the surface plasmon and evanescent coupling is a means to propagate light inside nanocircuits, the investigation of the coupled surface plasmons in the multilayer structures provides us with fundamental knowledge for the future 3D multilayer integration of the nanocircuits.  The structures considered here can be fabricated with existing nanotechnology.

This work is supported by ONR Independent Laboratory In-house Research funds.

\eject

\eject
\newpage

\bfg[h]
\centerline{\scalebox{.6}{\includegraphics{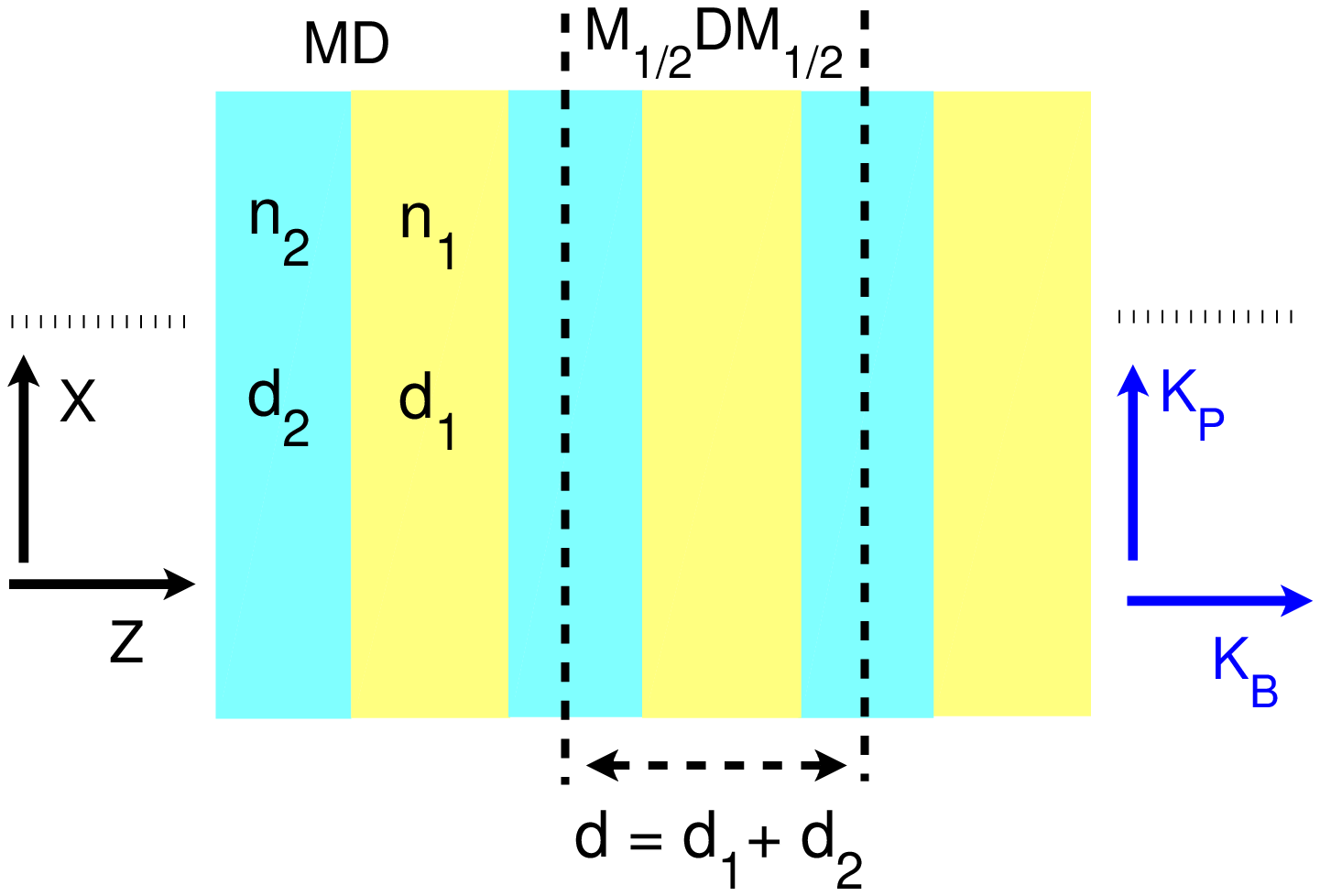}}}
\cpt{One-dimensional alternating layers of dielectric (subscript 1) and metallic materials (subscript 2).  The thicknesses of the dielectric and metal layers are $d_1$ and $d_2$, respectively, with period $d=d_1+d_2$.  This schematic shows two type unit cells, $MD$ and $M_{1/2}DM_{1/2}$.  The unit cell $M_{1/2}DM_{1/2}$ is defined by the section between two dashed lines in the middle of the metallic layers, so the unit cell possesses mirror symmetry whereas the unit cell $MD$ does not.  The surface plasmon wavevector is ${\bf K_p} = {\bf\hat x} K_p$ and the Bloch wavevector is ${\bf K_B} = {\bf\hat z} K_B$.  If the number of layers $N$ is infinite, the configuration of the unit cell would not be important, but for the finite number of layers, this is very important in terms of the transparent photonic band. In other words, the termination of the multilayer structure is important. The thicknesses of the layers are sufficiently thin such that the surface plasmon fields are evanescently coupled along the ${\bf\hat z}$ direction throughout the multilayer structure.   \label{StructCPRW}}
\efg

\bfg[h]
\centerline{\scalebox{.55}{\includegraphics{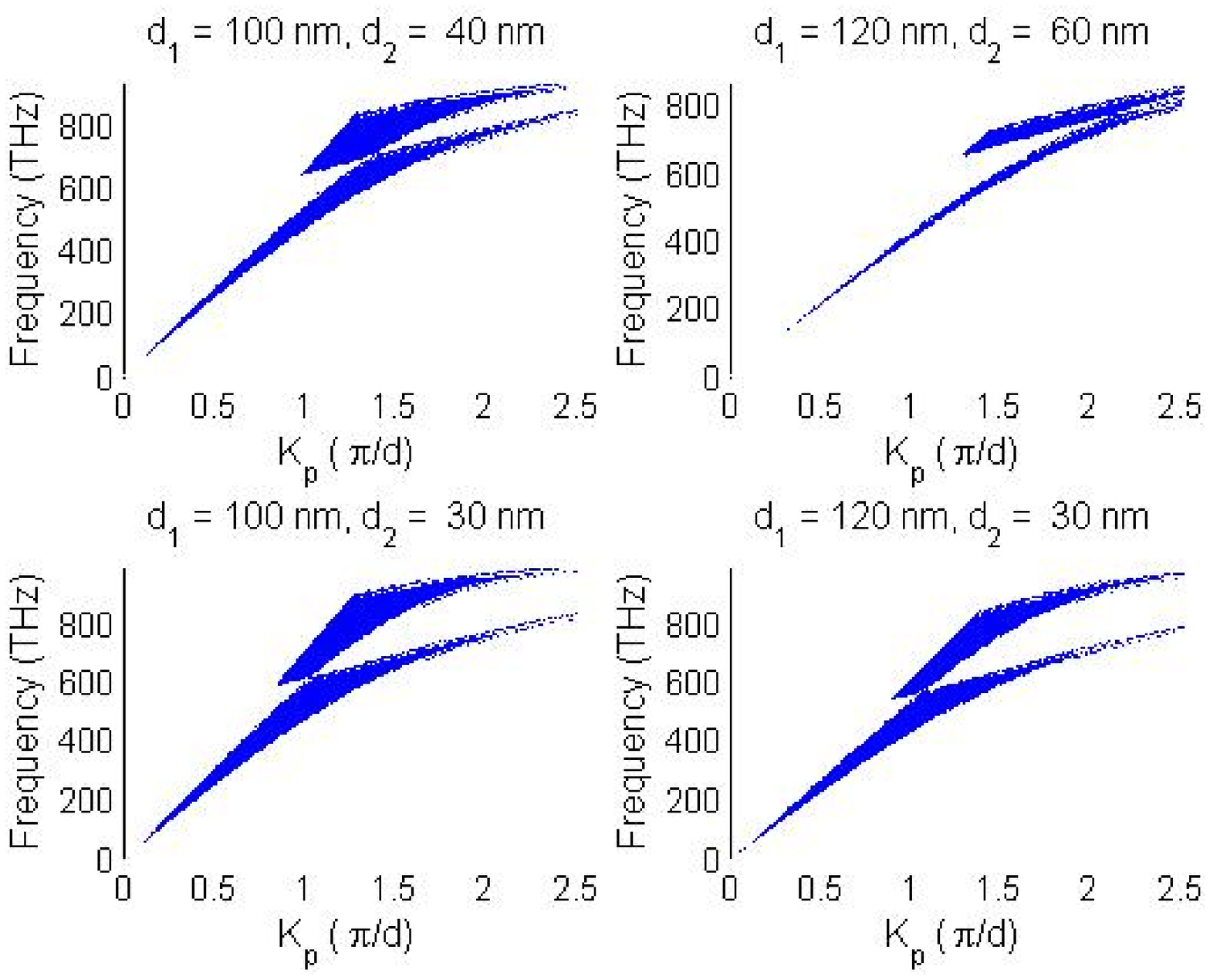}}}
\cpt{Band structure of TM mode evanescent fields.  The blue zones represent the pass bands of the evanescent waves.   These data were generated with $\epsilon_1 = 2.66$ and $\epsilon_2(\omega)=1-(\omega_p/\omega)^2$ with plasma frequency $\omega_p=1.2\times10^{16}\,\mbox{s}^{-1}$.  \label{HMband}}
\efg

\bfg[h]
\centerline{{\includegraphics[height=3in, width=3.8in]{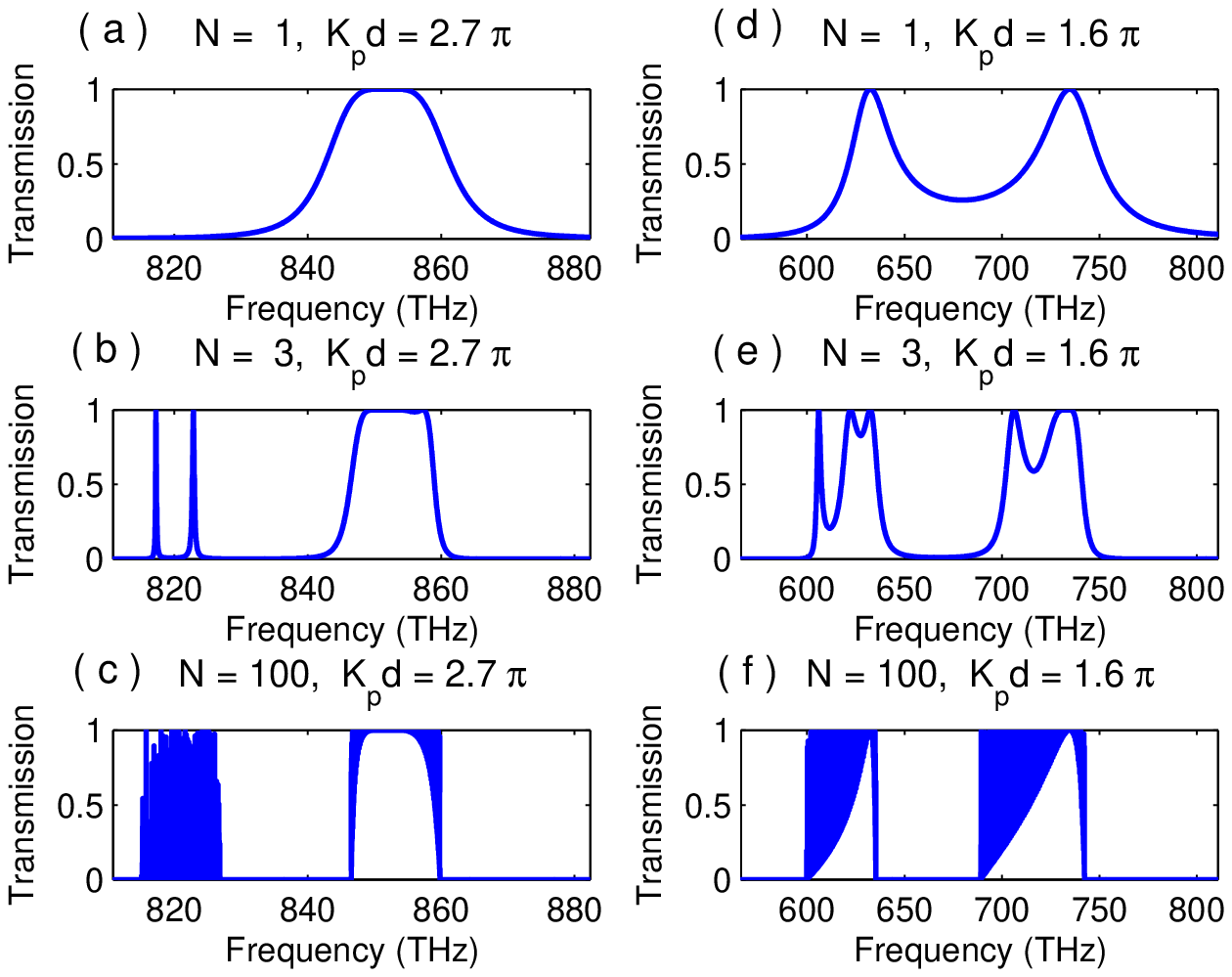}}}
\cpt{Transmission spectra showing evolution of the photonic bands with the number of periods ($N$) of the structure.  The unit cell is $M_{1/2}DM_{1/2}$ as defined in Fig.~\ref{StructCPRW}.  The layer thicknesses are $d_1=120$\,nm and $d_2=60$\,nm for all the plots and this corresponds to Fig.~\ref{HMband}(b).  Here, the plots (a), (b), and (c) have $K_p=2.7\pi/d$ and two pass bands.  The band centered around $850$ THz is a transparent band, and the band centered around 820 THz is not the transparent band.  The plots in (d), (e), and (f) have  $K_p=1.6\pi/d$ and neither pass band has the characteristics of the transparent band.  The important distinction between the transparent and non-transparent bands is that within the transparent band, the field intensity is symmetric in the unit cell.  Also, unlike the transparent band, the transmission in the non-transparent band oscillates rapidly with changing frequency.  \label{HMtransSpec}}
\efg

\bfg[h]
\centerline{\scalebox{.6}{\includegraphics{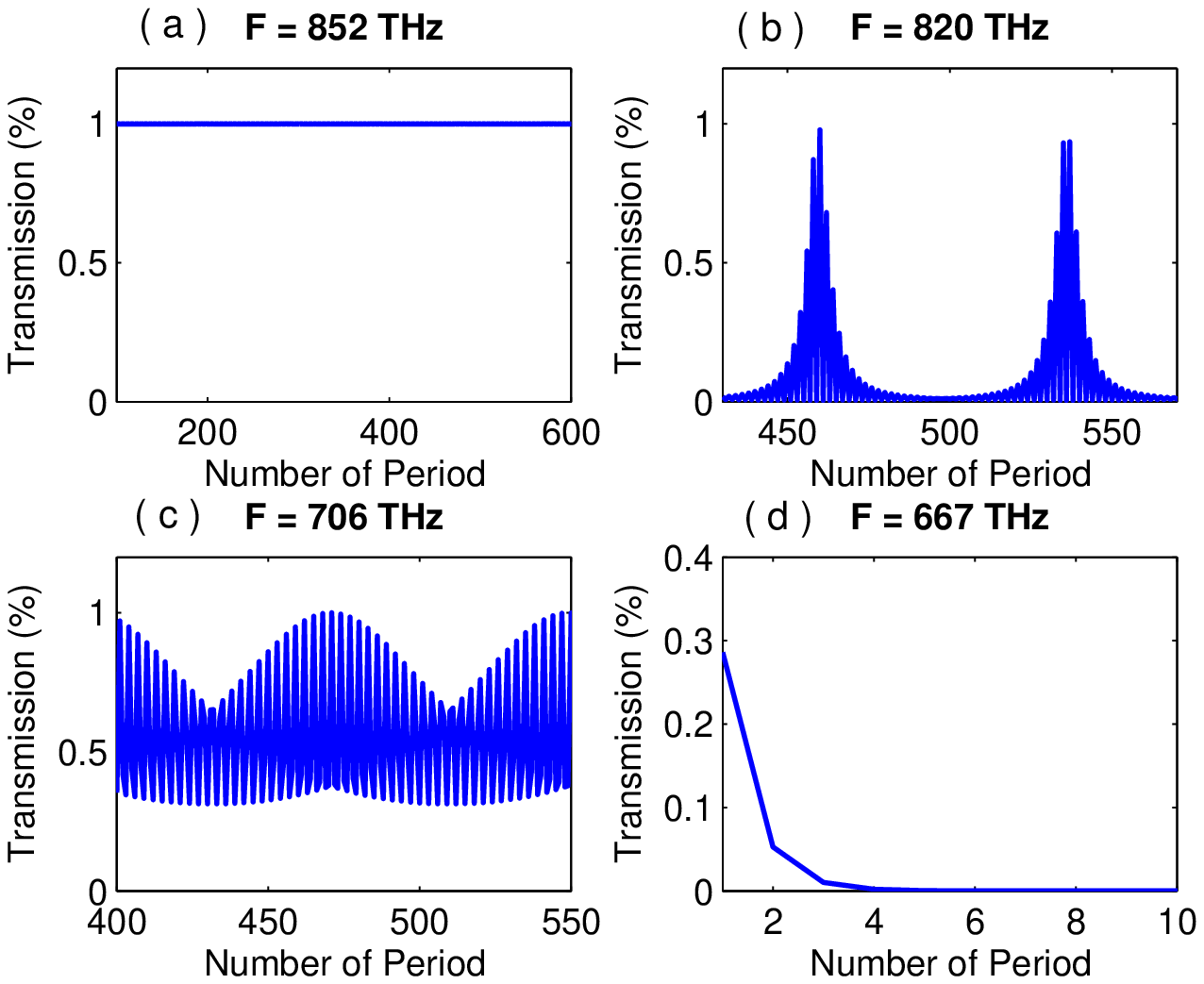}}}
\cpt{Four types of transmission coefficient characteristics versus the number of periods of the structure.  (a) Frequency $F=852$\,THz is in the higher frequency band (transparent band) in Fig.~\ref{HMtransSpec}(c).  (b) $F=820$\,THz is in the lower frequency band in Fig.~\ref{HMtransSpec}(c).  (c) $F=706$\,THz is in the pass band in Fig.~\ref{HMtransSpec}(f).  (d) $F=667$\,THz is in the stop band in Fig.~\ref{HMtransSpec}(f).  \label{HMtransCoef}}
\efg

\bfg[h]
\centerline{\scalebox{.6}{\includegraphics{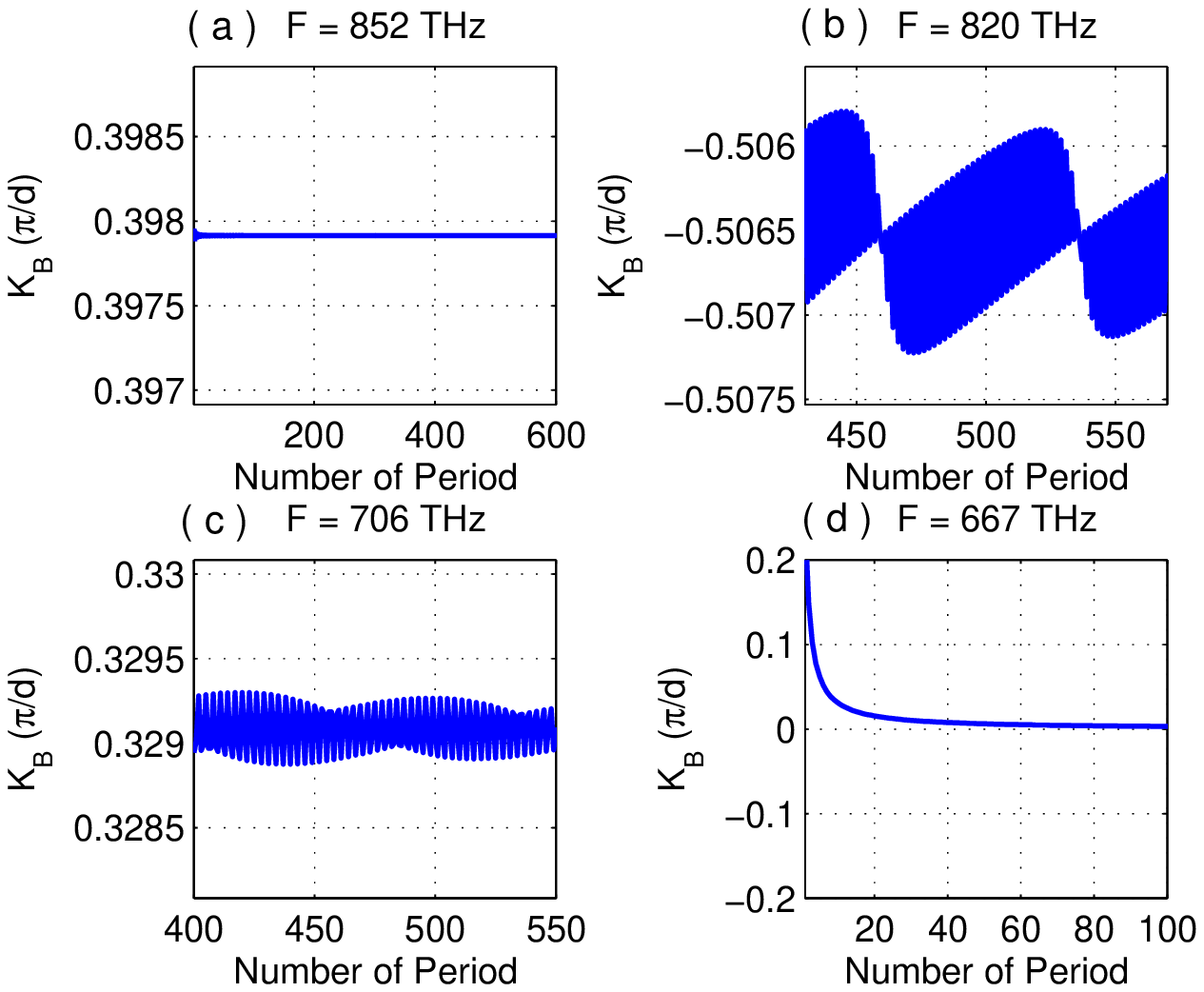}}}
\cpt{Bloch wavevector $K_B$ versus the number of periods of the structure for the frequencies in Fig.~\ref{HMtransCoef}.  \label{HMkvects}}
\efg

\bfg[h]
\centerline{\scalebox{.5}{\includegraphics{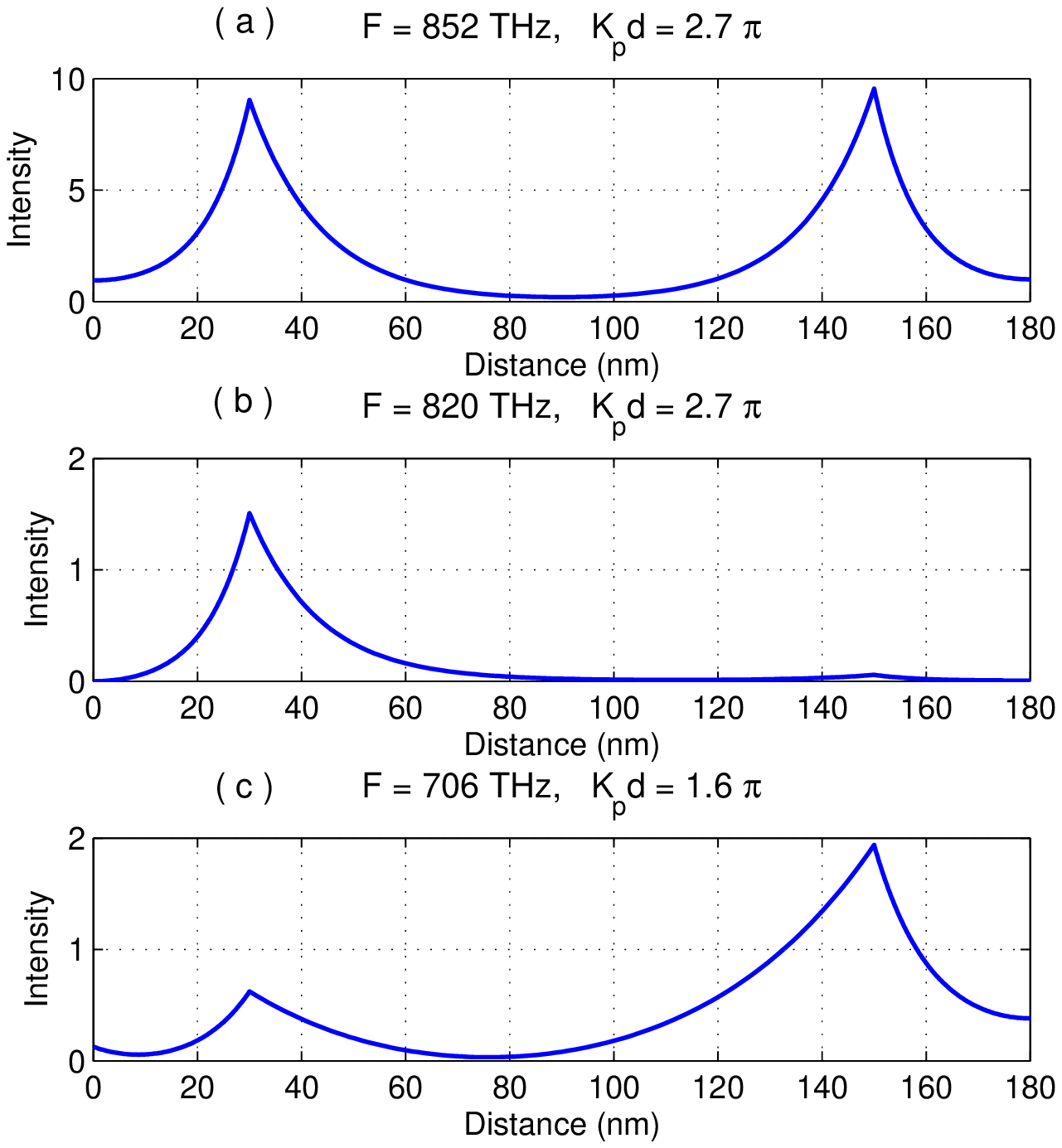}}}
\cpt{Intensity distribution within a single unit cell $M_{1/2}DM_{1/2}$ as defined in Fig.~\ref{StructCPRW}.  In (a)-(c) above, the parameters $F$ and $K_p$ can be correlated with those in Fig.~\ref{HMtransSpec} for $N=1$.  In (a) above, the intensity is symmetric and this is within the transparent band as seen from Fig.~\ref{HMtransSpec}(a).  In (b) and (c) above, the fields are not symmetric and are outside the transparent band.  \label{HMfield}}
\efg

\bfg[h]
\centerline{{\includegraphics[height=3in, width=3.8in]{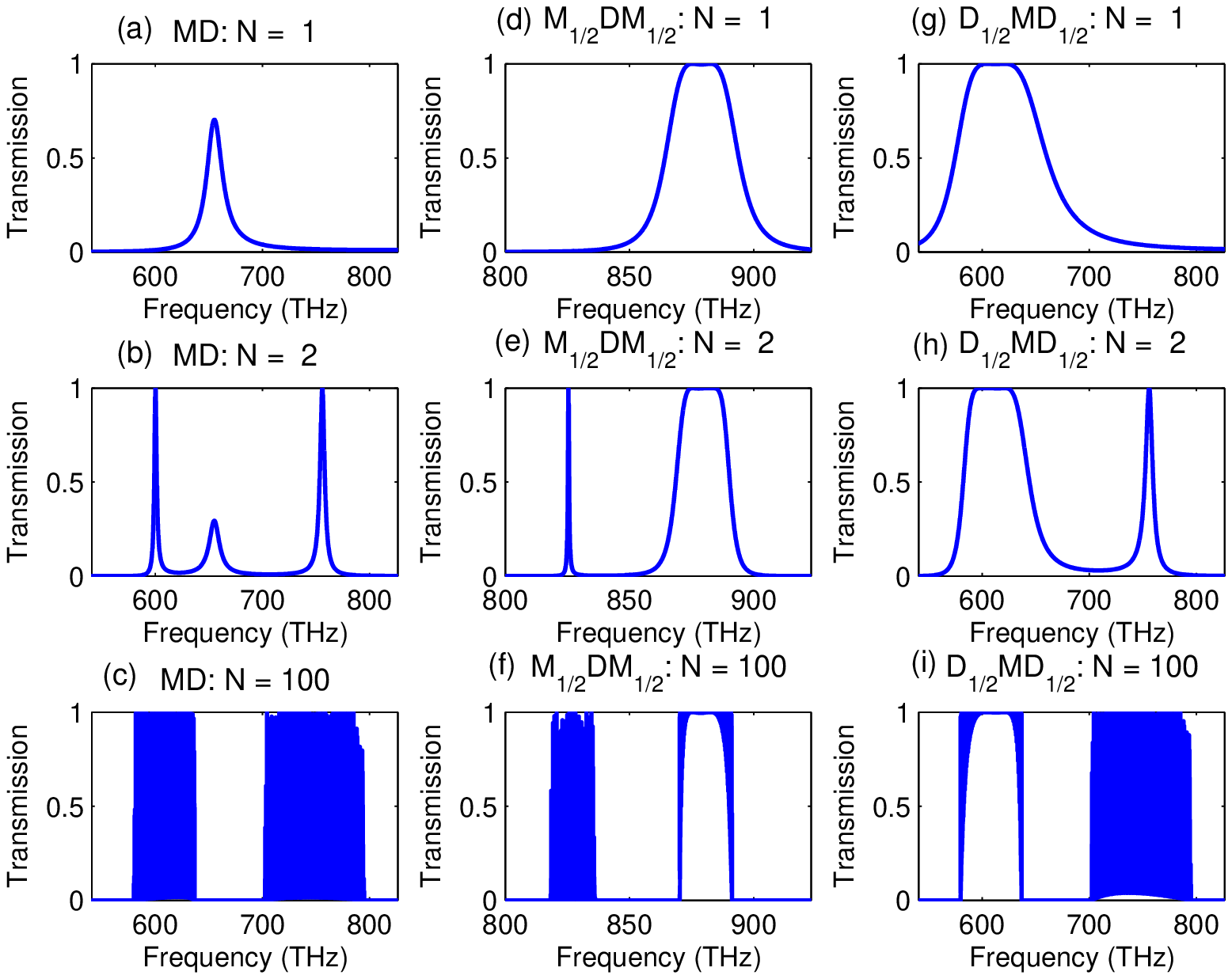}}}
\cpt{Transmission spectra showing evolution of the photonic bands with the number of periods $N$ for the three configurations $(MD)^N$, $(M_{1/2}DM_{1/2})^N$, and $(D_{1/2}MD_{1/2})^N$.  In all the plots, $d_1=100$\,nm and $d_2=50$\,nm.  The surface plasmon wavenumber $K_p=2.4\pi/d$ for $(M_{1/2}DM_{1/2})^N$ and $K_p=1.35\pi/d$ for $(D_{1/2}MD_{1/2})^N$ and $(MD)^N$.  Since the modes are different for different unit cells, the plasmon excitation condition is different to form the transparent band in the configurations $(M_{1/2}DM_{1/2})^N$ and $(D_{1/2}MD_{1/2})^N$. In other words, a transparent band exists for $(M_{1/2}DM_{1/2})^N$ ((d)-(f)) and $(D_{1/2}MD_{1/2})^N$ ((g)-(i)) although the surface plasmon mode ($K_p, F$) is different.  In the $(MD)^N$ case, the transparent band cannot be excited no matter what value of $K_p$. This is directly related to the fact that a symmetric surface plasmon mode cannot exist in the $MD$ unit cell.  \label{HDMtrans}}
\efg

\bfg[h]
\centerline{\scalebox{.6}{\includegraphics{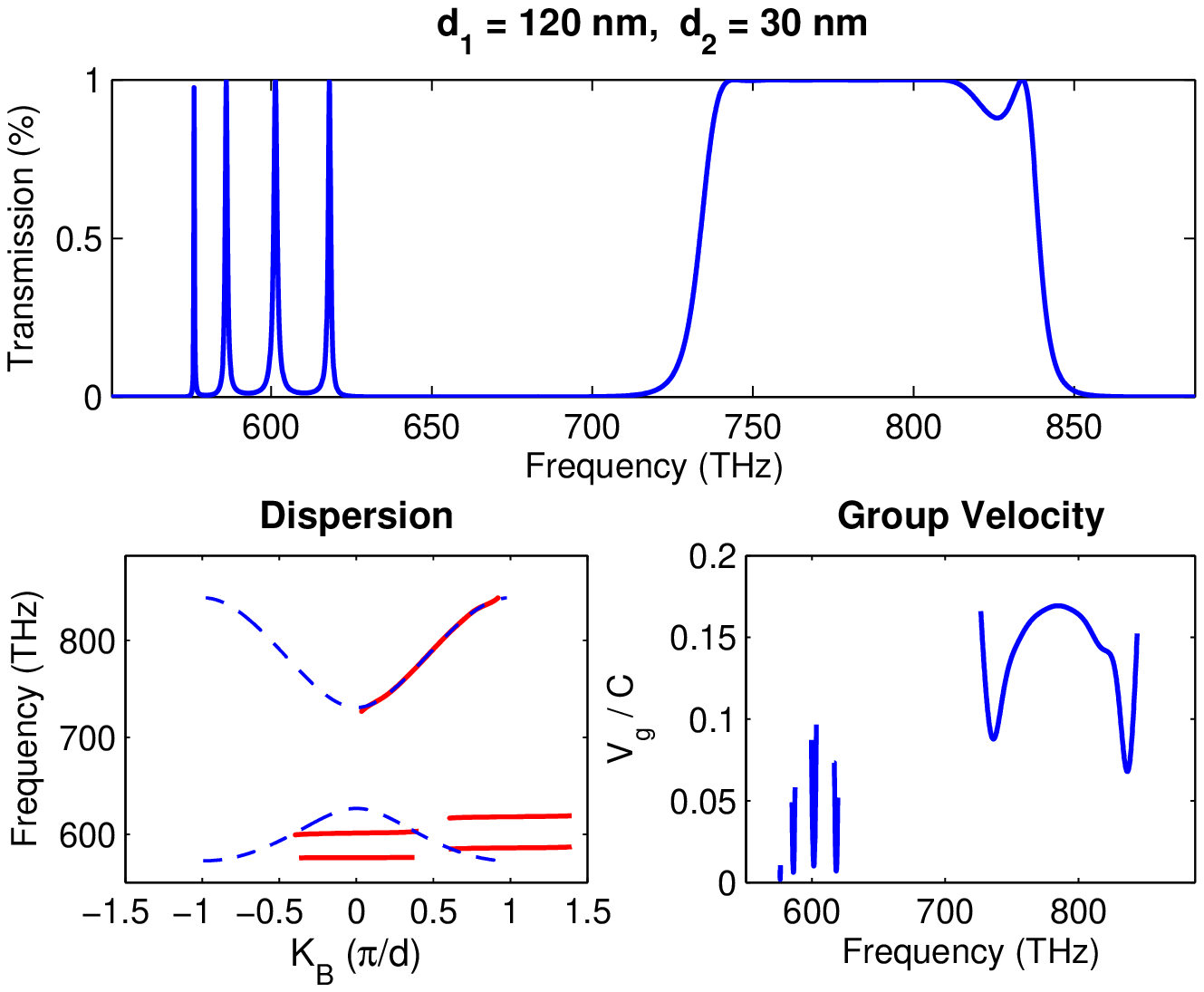}}}
\cpt{Top: transmission spectrum of a 5-period CPRW structure, i.e. $(M_{1/2}DM_{1/2})^5$, with $d_1=120$\,nm and $d_2=30$\,nm.  The plasma wavevector $K_p=1.45\pi/d$.  The bandwidth of the transparent band is about 100\,THz.  Bottom left: dispersion.  The red solid lines represent the dispersions of the transparent band (large slope) and the four resonances (small slope) in the lower band.  As a reference, the dashed blue curves show the dispersion of the corresponding infinite periodic structure.  Bottom right: group velocity $V_g$ relative to the speed of light in vacuum $c$.   \label{HMdisper}}
\efg


\newpage


\begin{thebibliography}{99}

\bibitem{Yablon}  E. Yablonovitch,  Phys. Rev. Lett. {\bf58}, 2059 (1987).

\bibitem{Ho}  K. M. HO, C. T. Chan, and C. M. Soukoulis,  Phys. Rev. Lett. {\bf65}, 3152-3155 (1990).

\bibitem{Bendickson}  J. M. Bendickson, J. P. Dowling, and M. Scalora,  Phys. Rev. E {\bf53}, 4107 (1996).

\bibitem{Olivier}  S. Olivier, C. Smith, M. Rattier, H. Benisty, C. Weisbuch, T. Krauss, R. Houdr{\'e}, and U. Oesterl{\'e},  Opt. Lett. {\bf26}, 1019 (2001).

\bibitem{Ye}  Y.-H. Ye, J. Ding, D.-Y. Jeong, I. C. Khoo, and Q. M. Zhang,  Phys. Rev. E {\bf69}, 056604 (2004).

\bibitem{Robertson}  W. M. Robertson,  J. Lightwave Technol. {\bf17}, 2013 (1999).

\bibitem{Elson}  J. M. Elson and K. Halterman,  Opt. Express {\bf12}, 4855 (2004).

\bibitem{Scalora}  M. Scalora, M. J. Bloemer, A. S. Pethel, J. P. Dowling, C. M. Bowden, and A. S. Manka,  J. Appl. Phys. {\bf83}, 2377-2383 (1998).

\bibitem{Yariv}  A. Yariv, Y. Xu, R. K. Lee, and A. Scherer,  Opt. Lett. {\bf24}, 711 (1999).

\bibitem{Kuzmiak}  V. Kuzmiak, A. A. Maradudin, and F. Pincemin,  Phys. Rev. B {\bf 50}, 16835-16844 (1994).

\bibitem{Sigalas}  M. M. Sigalas, C. T. Chan, K. M. Ho, and C. M. Soukoulis,  Phys. Rev. B {\bf 52}, 11744-11751 (1995).

\bibitem{Fan}  S. Fan, P. R. Villeneuve, and J. D. Joannopoulos,  Phys. Rev. B {\bf54}, 11245 (1996).

\bibitem{Simin}  For convenience, here the unit cell $MD$ is used.  The definition of the unit cell does not affect the band calculation because the photonic bands are computed for the infinite structure.  For the infinite structure, the three configurations $(MD)^\infty$, $(M_{1/2}DM_{1/2})^\infty$, and $(D_{1/2}MD_{1/2})^\infty$ should give the same result. Indeed, this is confirmed by our calculation.

\bibitem{Yanik}  M. F. Yanik and S. Fan,  Phys. Rev. A {\bf71}, 013803 (2005).

\end{thebibliography}
\end{document}